\documentstyle[12pt,epsfig]{article}

\begin{document}

\begin{titlepage}

\renewcommand{\thefootnote}{\fnsymbol{footnote}}


\vspace{0.3cm}

\begin{center}
{\Large\bf Strong field lensing by Damour-Solodukhin wormhole}
\end{center}

\begin{center}
K.K. Nandi\footnote{Electronic address: kamalnandi1952@rediffmail.com}$^{1,2}$, 
R.N. Izmailov\footnote{Electronic address: izmailov.ramil@gmail.com}$^{1}$,
E.R. Zhdanov\footnote{Electronic address: zhdanov@ufanet.ru}$^{1}$, 
Amrita Bhattacharya$\footnote{Electronic address: amrita\_852003@yahoo.co.in}^{3}$\par
\end{center}

\begin{center}

${}^{1}$Zel'dovich International Center for Astrophysics, Bashkir State Pedagogical University, 3A, October Revolution Street, Ufa 450008, RB, Russia, \\
${}^{2}$High Energy and Cosmic Ray Research Center, University of North Bengal, Siliguri 734 013, WB, India, \\
${}^{3}$Department of Mathematics, Kidderpore College, 2, Pitamber Sircar Lane, Kolkata 700023, WB, India \\
\end{center}

\vfill

\begin{abstract}

We investigate the strong field lensing observables for the Damour-Soludukhin wormhole and examine how small the values of the deviation parameter $\lambda $ need be for reproducing the observables for the Schwarzschild black hole. While the extremely tiny values of $\lambda$ indicated by the matter accretion or Hawking evaporation are quite consistent with the lensing observations, it turns out that $\lambda $ could actually assume values considerably higher values and still reproduce black hole lensing signatures. The lensing observables for SgrA* can be interpreted to provide an upper bound on $\lambda \sim 10^{-3}$ and until lower bound is established, all values of $\lambda $ below the upper bound should be treated as equally probable.
\end{abstract}

\vskip20pt

\end{titlepage}


\section{Introduction}
\label{sec:int}

Damour and Solodukhin (DS) [1] defined black hole "foils" as objects that mimic some aspects of black holes, while differ in other aspects. An ingenious toy model of such an object is what we shall call here the DS wormhole. Wormholes are solutions of Einstein's and other theories of gravity sourced by exotic matter (matter violating known energy conditions) although objects have not yet been ruled out by experiments. On the contrary, there are useful applications of Ellis wormholes as galactic halo objects [2]. A fundamental theoretical distinction between a black hole and a wormhole is that while the former possesses event horizon, the latter does not. Despite this distinction, it is found that many strong field features previously thought of as indicative of a black hole event horizon (e.g., ring-down quasi-normal modes) can be remarkably mimicked by a static wormhole [3-7].

A very effective tool for sampling strong field regime of gravity is provided by the gravitational lensing phenomenon, be it by a BH [8] or a wormhole [9], when the light rays pass arbitrarily close to the photon sphere in either case. Light rays passing infinitesimally close to the photon sphere wind up a large number of times before escaping away. While the the theoretical effect of light bending plays the role of core physics (see, e.g., [10,11]), the observable effect of gravitational lensing is a step ahead providing observable set of values that may constitute a kind of "identity card"\ (name coined by Bozza [12]) for different types of lenses. Since good evidences exist in favor of black holes (e.g., each galaxy is believed to host a black hole in its center), a curious question is to what extent a wormhole can reproduce strong field lensing observables of a Schwarzschild black hole.

DS wormhole introduces a deviation parameter $\lambda $ in the Schwarzschild metric converting it into a black hole foil. The event horizon then is replaced by a high tension distribution of exotic matter localized around the wormhole throat at $r_{\textmd{th}}=2GM$. DS found that many observational features of a Schwarzschild black hole on classical and quantum level could be well mimicked by a wormhole, if the parameter $\lambda $ is sufficiently (exponentially) small, $\lambda \sim e^{-4\pi GM^{2}}$. They argued that the only way to observationally distinguish a wormhole from a black hole is to observe the classical effect of matter accretion over the long \textquotedblleft wormhole bounce\textquotedblright\ time scale $\Delta t=2GM\ln \left(\frac{1}{\lambda^{2}}\right)$, or quantum effect of Hawking evaporation again over a long time scale $\Delta t=16\pi G^{2}M^{3}$. Further, the accretion effect shows that if $\lambda $ is small enough, it is impossible for observations over some limited time interval $\Delta T$ to distinguish the fall of matter onto the throat of a wormhole from the fall into the horizon of a black hole since $\Delta t\gg \Delta T$. As an example, assuming that the candidate black hole SgrA* in our galaxy started accreting matter $6$ billion years ago, it could be a wormhole if $\lambda\ll e^{-10^{15}}$, an incredibly tiny value indeed [1].

The purpose of the present paper is to investigate the strong field lensing observables for the DS wormhole and examine how small the values of the deviation parameter $\lambda $ need be for reproducing the observables for the Schwarzschild black hole. We shall use Bozza's method [8] for calculating the observables and provide an upper bound on $\lambda$.

The paper is organized as follows. In Sec.2, an outline of Bozza's method is given and in Sec.3 it is applied to DS wormhole. Numerical comparisons are presented in Sec.4, while Sec.5 concludes the paper. We take units such that $8\pi G=1$, $c=1$ unless specifically restored.

\section{Bozza's method in outline}
\label{sec:met}

This method has by now gained considerable attention for its usefulness. The purpose of this preview is to ensure clarity by letting the readers readily see what quantities have been calculated to get to the final lensing observables. The method starts with a generic spherically symmetric static spacetime
\begin{equation}
ds^{2}=A(x)dt^{2}-B(x)dx^{2}-C(x)\left( d\theta ^{2}+\sin ^{2}\theta \phi^{2}\right).
\end{equation}%
The equation [16,17]
\begin{equation}
\frac{C^{\prime}(x)}{C(x)}=\frac{A^{\prime}(x)}{A(x)}
\end{equation}%
is assumed to admit at least one positive root and the largest root is called the radius of the photon sphere $x_{m}$ (the subscript $m$ meaning minimum radius). The strong field expansion will take the photon sphere radius as the starting point, which is required to exceed the horizon radius of a black hole or throat radius of a wormhole as the case may be. A light ray coming in from infinity will reach the closest approach distance $x_{0}$ from the centre of the gravitating source before emerging in another direction. By the conservation of angular momentum, $x_{0}$ is related to the impact parameter $u$ by
\begin{equation}
u=\sqrt{\frac{C_{0}}{A_{0}}}
\end{equation}%
where the subscript $0$ indicates that the function is evaluated at $x_{0}$. The minimum impact parameter is defined by%
\begin{equation}
u_{m}=\sqrt{\frac{C_{m}}{A_{m}}},
\end{equation}%
where $C_{m}\equiv C(x_{m})$ etc. From the null geodesics, the deflection angle $\alpha (x_{0})$ can then be calculated as a function of the closest approach [18]:
\begin{eqnarray}
\alpha (x_{0}) &=& I(x_{0})-\pi, \\
I(x_{0}) &=&\int\limits_{x_{0}}^{\infty }\frac{2\sqrt{B}dx}{\sqrt{C}\sqrt{\frac{%
C}{C_{0}}\frac{A_{0}}{A}-1}}.
\end{eqnarray}

In the weak field limit of deflection, the integrand is expanded to any order in the gravitational potential and integrated. When we decrease the impact parameter $u$ (and consequently $x_{0}$), the deflection angle increases. Decreasing $u$ further bringing the ray infinitesimally closer to the photon sphere will cause the ray to wind up a large number of times before emerging out. Finally, at $x_{0}=x_{m}$, corresponding to an impact parameter $u=u_{m}$, the deflection angle will diverge and the ray will be captured, i.e., it will wind around the photon sphere indefinitely.

Bozza [8] has shown that this divergence is logarithmic for all spherically symmetric metrics, which yields an analytical expansion for the deflection angle close to the divergence in the form
\begin{equation}
\alpha (x_{0})= -a\log\left(\frac{x_{0}}{x_{m}}-1\right) +b+O\left(x_{0}-x_{m}\right).
\end{equation}%
The coefficients $a,b$ depend on the metric functions evaluated at $x_{m}$, and Eq.(2.7) defines the \textit{strong field limit} $x_{0}\rightarrow x_{m}$ of the light deflection angle. Next define two new variables
\begin{eqnarray}
y &=& A(x), \\
z &=&\frac{y-y_{0}}{1-y_{0}},
\end{eqnarray}
where $y_{0}=A_{0}$. The integral (2.6) then becomes
\begin{eqnarray}
I(x_{0}) &=& \int\limits_{0}^{1}R(z,x_{0})f(z,x_{0})dz, \\
R(z,x_{0}) &=& \frac{2\sqrt{By}}{CA^{\prime }}\left( 1-y_{0}\right) \sqrt{C_{0}}, \\
f(z,x_{0}) &=& \frac{1}{\sqrt{y_{0}-\left[ \left( 1-y_{0}\right) z+y_{0}\right]\frac{C_{0}}{C}}},
\end{eqnarray}
where all functions without the subscript $0$ are evaluated at $x=A^{-1}\left[ \left( 1-y_{0}\right) z+y_{0}\right] $. The function $R(z,x_{0})$ is regular for all values of $z$ and $x_{0}$, while $f(z,x_{0})$ diverges for $z\rightarrow 0$, where
\begin{eqnarray}
f(z,x_{0}) &\sim & f_{0}(z,x_{0})=\frac{1}{\sqrt{\alpha z+\beta z^{2}}}, \\
\alpha &=& \frac{1-y_{0}}{C_{0}A_{0}^{\prime }}\left( C_{0}^{\prime}y_{0}-C_{0}A_{0}^{\prime }\right), \\
\beta &=& \frac{\left( 1-y_{0}\right) ^{2}}{2C_{0}^{2}{A_{0}^{\prime }}^{3}}\left[ 2C_{0}C_{0}^{\prime }{A_{0}^{\prime }}^{2}+\left( C_{0}C_{0}^{\prime \prime }-2{C_{0}^{\prime }}^{2}\right) y_{0}A_{0}^{\prime }-C_{0}C_{0}^{\prime }y_{0}A_{0}^{\prime \prime }\right],
\end{eqnarray}
where primes denote differentiation with respect to $x$.

For the calculation of lensing observables, note that the angular separation of the image from the lens is $\tan \theta =\frac{u}{D_{\textmd{OL}}}$, where $D_{\textmd{OL}}$ is the distance between the observer and the lens [8]. Specializing to the photon sphere $x_{0}=x_{m}$, the deflection angle in Eq.(2.7) can be rewritten into a final form

\begin{eqnarray}
\alpha (\theta ) &=&-\overline{a}\log \left( \frac{u}{u_{m}}-1\right) +\overline{b}, \\
u &\simeq &\theta D_{\mathrm{OL}}\textmd{ (assuming small }\theta \textmd{)}, \\
\overline{a} &=& \frac{a}{2}=\frac{R(0,x_{m})}{2\sqrt{\beta _{m}}}\textmd{, } \\
\overline{b} &=&-\pi +b_{R}+\overline{a}\log {\frac{2\beta _{m}}{y_{m}}}\textmd{, } \\
y_{m} &=&A(x_{m}), \\
\beta _{m} &=&\beta |_{x_{0}=x_{m}}\textmd{,} \\
b_{R} &=&\int_{0}^{1}g(z,x_{m})dz, \\
g(z,x_{m}) &=&R(z,x_{m})f(z,x_{m})-R(0,x_{m})f_{0}(z,x_{m}).
\end{eqnarray}

The impact parameter $u$ is related to the angular separation $\theta $ of
images by the relationship given in Eq.(2.17). Using this, Bozza proposed
three strong lensing observables as [8]
\begin{eqnarray}
\theta _{\infty } &=&\frac{u_{m}}{D_{\textmd{OL}}}, \\
s &=&\theta _{\infty }\exp \left( \frac{\bar{b}}{\bar{a}}-\frac{2\pi }{\bar{a%
}}\right) , \\[0.5em]
r &=&2.5\log _{10}\left[ \ \exp \left( \frac{2\pi }{\bar{a}}\right) \ \right],
\end{eqnarray}%
where $\theta _{\infty }$ is the asymptotic position approached by a set of
images in the limit of a large number of loops the rays make around the
photon sphere ( $\theta _{\infty }$ ia also called the angular radius of the
black hole\textit{\ shadow} [13]), $s$ is the angular separation between the
outermost image resolved as a single image and the set of other asymptotic
images, all packed together, while $r$ is ratio between the flux of the
first image and the flux coming from all the other images.

We shall calculate in the next section the strong field lensing coefficients
$\left\{ \overline{a},\overline{b},u_{m}\right\} \ $and the resultant
observables $\left( \theta _{\infty },s,r\right) $ applying respectively the
formulas (2.18)-(2.19) and (2.22)-(2.24) to the DS wormhole. The set $%
\left\{ \overline{a},\overline{b},u_{m}\right\} $ defines the "identity
card" of the concerned lens that differs from lens to lens.

\section{Application to DS wormhole}
\label{sec:app}

For future works, it will be useful to have the relevant expressions in one
place, some of which are:
\begin{eqnarray}
R(z,x_{0}) &=& \left( \frac{2M-\lambda ^{2}x_{0}}{M}\right) \left( \sqrt{\frac{1+2z-z\lambda ^{2}+\lambda ^{2}}{1+2z-z\lambda ^{2}-2\lambda ^{2}}}\right), \\
f(z,x_{0}) &\sim & f_{0}(z,x_{0})=\frac{1}{\sqrt{\alpha z+\beta z^{2}}}, \\
\alpha &=& \frac{\left( \lambda ^{2}x_{0}-2M\right) \left\{ 3M-x_{0}(1+\lambda^{2})\right\} }{Mx_{0}}, \\
\beta &=& \frac{\left(\lambda^{2}x_{0}-2M\right)^{2}\left\{6M-x_{0}(1+\lambda ^{2})\right\} }{4M^{2}x_{0}}
\end{eqnarray}
The radius $x_{m}$ of the photon sphere can be found from Eq.(2.2) as (it also
follows from $\alpha =0$)%
\begin{equation}
x_{m}=\frac{3M}{1+\lambda ^{2}},
\end{equation}%
which yields the expressions
\begin{eqnarray}
R(z,x_{m}) &=& \left( \frac{2-\lambda ^{2}}{1+\lambda ^{2}}\right) \left( \sqrt{\frac{1+2z-z\lambda ^{2}+\lambda ^{2}}{1+2z-z\lambda ^{2}-2\lambda ^{2}}}\right), \nonumber\\
\alpha _{m} &=& \alpha |_{x_{0}=x_{m}}=0,\beta _{m}=\beta |_{x_{0}=x_{m}}=\frac{%
\left( \lambda ^{2}-2\right) ^{2}}{4(1+\lambda ^{2})},\textmd{ }y_{m}=A(x_{m})=%
\frac{1+\lambda ^{2}}{3}, \\
g\left( z,x_{m}\right) &=& \frac{2}{z}\left[ \sqrt{\frac{3\left\{ (1+\lambda
^{2})-z(\lambda ^{2}-2)\right\} }{\left\{ (1-2\lambda ^{2})-z(\lambda
^{2}-2)\right\} \left\{ z(\lambda ^{2}-2)+3(1+\lambda ^{2})\right\} }}-\frac{%
1}{\sqrt{1-2\lambda ^{2}}}\right].
\end{eqnarray}
and the minimum impact parameter $u_{m}$ follows from Eq.(2.4) as
\begin{equation}
u_{m}=3\sqrt{3}M(1+\lambda ^{2})^{3/2}.
\end{equation}%
Then it follows from Eqs.(2.18)-(2.19) that the exact coefficients are
\begin{eqnarray}
\overline{a} &=&\frac{1}{\sqrt{1-2\lambda ^{2}}} \\
\overline{b} &=&-\pi +b_{R}+\overline{a}\log \left[ \frac{3}{2}\frac{\left(
\lambda ^{2}-2\right) ^{2}}{\left( 1+\lambda ^{2}\right) ^{2}}\right] , \\
b_{R}(\lambda ) &=&\int_{0}^{1}g(z,x_{m})dz, \\
g\left( z,x_{m}\right) &=&\frac{2}{z}\left[ \sqrt{\frac{3\left\{ (1+\lambda
^{2})-z(\lambda ^{2}-2)\right\} }{\left\{ (1-2\lambda ^{2})-z(\lambda
^{2}-2)\right\} \left\{ z(\lambda ^{2}-2)+3(1+\lambda ^{2})\right\} }}-\frac{%
1}{\sqrt{1-2\lambda ^{2}}}\right] . \nonumber
\end{eqnarray}

\section{Numerical comparisons}
\label{sec:num}

For numerical illustration of the strong lensing signatures, we choose as a
candidate massive black hole SgrA* residing at our galactic center\footnote{%
Strictly speaking, SgrA* black hole has spin and for comparison of
observables, the appropriate wormhole examples should also be the spinning
ones. While choosing SgrA* is not mandatory for our analysis, we point out
that SgrA* has been modeled purely as a Schwarzschild black hole in the
literature [13].}. The table below shows how the variation of $\lambda $ leads to
variations in the strong field lensing observables, when SgrA* is portrayed
as a DS\ wormhole $\left(\lambda \neq 0\right) $ and as a Schwarzschild
black hole $\left( \lambda =0\right) $. We take the SgrA* values to be: $%
M=4\times 10^{6}M_{\odot }$, $D_{\textmd{OL}}=8$ kpc [15], $r$ (magnitude)$%
=2.5\times $ log$_{10}\left( r\right) .$

\begin{table}[!ht]
\centering
\begin{tabular}{|c|c|c|c|c|c|c|c|}
\hline
Lens & $\lambda $ & $\overline{a}$ & $\overline{b}$ & $u_{m}\times
10^{12}$ cm & $\theta _{\infty }$($\mu$as) & $s$($\mu $as) & $r$(mag) \\
\hline
 & $0.05$ & $1.0025$ & $-0.4163$ & $3.0876$ & $25.8054$ & $0.0323$ & $6.8048$ \\
 & $0.04$ & $1.0016$ & $-0.4105$ & $3.0835$ & $25.7707$ & $0.0322$ & $6.8109$ \\
DS wh & $0.03$ & $1.0009$ & $-0.4046$ & $3.0802$ & $25.7436$ & $0.0322$ & $6.8157$ \\
 & $0.02$ & $1.0004$ & $-0.4028$ & $3.0779$ & $25.7244$ & $0.0321$ & $6.8191$ \\
 & $0.01$ & $1.0001$ & $-0.4008$ & $3.0765$ & $25.7128$ & $0.0321$ & $6.8212$ \\
 & $0.001$ & $1.0000$ & $-0.4002$ & $3.0761$ & $25.7089$ & $0.0321$ & $6.8218$ \\
\hline
Sch bh & $0$ & $1.0000$ & $-0.4002$ & $3.0761$ & $25.7089$ & $0.0321$ & $ 6.8218$ \\
\hline
\end{tabular}
\caption{\label{tab:1} Strong field lensing coefficients and observables for DS wormhole.}
\end{table}

\section{Summary}
\label{sec:sum}

It is evident from the table that the strong field coefficients ($\overline{a},\overline{b}$) and observables ($u_{m}$,$\theta _{\infty },s,r$) are very close to those of Schwarzschild when $0.001\leq \lambda \leq 0.05$. For $\lambda \leq 0.001$, the values coincide with those of Schwarzschild black hole up to any accuracy. As the last two rows show, for $\lambda \sim 10^{-3} $, lensing coefficients and observables\ exactly reproduce known Schwarzschild values accurate up to four decimal places [8]. Among the lensing observables, the best one from the observational point of view is the angular diameter of the shadow $2\theta _{\infty }$ cast upon the background accretion flow. To measure the diameter for SgrA* shadow, the Event Horizon Telescope (EHT) array have been set up with international collaboration. The EHT is primarily expected to achieve a resolution of only $15$ $\mu $as at $345$ GHz [14] against the basic value $\sim 50$ $\mu $as (see Table). The first set of EHT data is expected to be available approximately in July 2018. However, it is evident that the experimental sensitivity needed to distinguish between the DS wormhole and black hole is still far ahead in the future.

The range $\lambda \leq 10^{-3}$ from strong field lensing is quite
compatible with tiny values suggested by other observations in [1].
Therefore, at best $\lambda \sim 10^{-3}$ can be interpreted as an upper
bound on the parameter. What however is more important is to constrain $%
\lambda $ from below. Plausible lower bounds appear to come from the example
of matter accretion or Hawking evaporation or some other experiment. Hence,
we conclude that until a lower bound is established, all values of $\lambda $
below the upper bound should be treated as equally probable. A similar, but
not the same, type of categorization of compact object radii differing from
those of black holes by a parameter $\varepsilon $ have been found by
Cardoso and Pani [5] in the context of the emission of gravitational waves.

\section*{Acknowledgement}

We thank the anonymous referee for helpful comments. The reported study was
funded by RFBR according to the research project No. 18-32-00377.


\bibliography{references}

\end{document}